\documentclass[prb,twocolumn,superscriptaddress,floatfix]{revtex4}
\usepackage{graphics}
\usepackage{graphicx}
\usepackage{float}
\usepackage{amsfonts}
\usepackage{amssymb}
\usepackage{textcomp}
\usepackage{color}
\def\s3s3{$\sqrt{3}\times\sqrt{3}$}

\usepackage{epsfig}
\usepackage{amsmath,hyperref}
\usepackage{wasysym}
\usepackage{latexsym}

\newcommand{\be}{\begin{eqnarray}}
\newcommand{\ee}{\end{eqnarray}}

\def\unue#1{{\it#1:}}

\begin{document}

\title{A topological spin glass in diluted spin ice}

\author{Arnab Sen}
\affiliation{Department of Theoretical Physics, Indian Association for the Cultivation of Science, Kolkata-700032, India}
\affiliation{Max-Planck-Institut f\"ur Physik komplexer Systeme,  01187 Dresden, Germany}
\author{R. Moessner}
\affiliation{Max-Planck-Institut f\"ur Physik komplexer Systeme,  01187 Dresden, Germany}

\date{\today}

\begin{abstract}
It is a salient experimental fact that a large 
fraction of candidate spin liquid materials freeze as the temperature is lowered. 
The question naturally arises whether such freezing is  intrinsic to the spin liquid (``disorder-free glassiness'') 
or extrinsic, in the sense that a topological phase simply coexists with standard freezing of impurities. 
Here, we demonstrate a surprising third alternative, namely that freezing and topological liquidity are inseparably linked.
The topological phase reacts to the introduction of disorder by generating degrees of freedom of a new type (along with 
interactions between them), which in turn undergo a 
freezing transition while the topological phase supporting them remains intact. 
\end{abstract}


\maketitle

\unue{Topology, liquidity and glassiness} The search for topological states of matter in magnetism 
over the last two decades has produced a good number of candidate classical and quantum spin 
liquids \cite{PALee} which show no conventional ordering down to temperatures much lower than the 
energy scale $\Theta_W$ of their interactions. 
In experiment, such behaviour often goes along with 
 glassy behavior, be it for Ising \cite{spiniceglass} or Heisenberg 
\cite{SCGO,GGG, quasisg} spins; 
in dimensions $d = 2$ \cite{SCGO, jarosites, volborthite} or $d = 3$ \cite{spiniceglass,GGG}; and for different
disorder types, e.g.\ distortions\cite{ymol} or dilution \cite{quasisg}. In this way, the 
study of glassy physics has become one of the staples of the field. A comprehensive discussion is provided in 
Ref.~\onlinecite{cepascanals}.

A systematic understanding of the rich experimental findings has been slow to emerge. Even the minimal 
ingredients to obtain freezing remain unclear. A seductive idea was the prospect of disorder-free glassiness, where a 
rugged energy landscape was posited to exist {\em even in the absence of quenched disorder}, thus accounting 
for the slow dynamics \cite{PIFrance3_1993,cepascanals}. A more pedestrian alternative is the realist view that 
any system exhibits some level of quenched disorder, and hence a tendency towards glassy behaviour, which 
becomes frequently visible in the case of spin liquids as there, it is not preempted by other instabilities. Indeed, 
in the case of bond disorder, \cite{CaGiHoMo_2001} the existence of a conventional glass transition in the 
pyrochlores was shown to occur at a critical temperature set simply by the amplitude of the bond disorder.\cite{chalker}

Our work presents a third way towards glassiness in topological spin states: the interplay of disorder with the 
topological phase produces {\it emergent} degrees of freedom along with interactions between them; 
it is these new degrees of freedom which in turn undergo a freezing transition. 
This combination of ``extrinsic'' disorder teaming up with
 ``intrinsic'' 
properties of the topological state presents an attractive conceptual angle on the {ubiquity} of spin freezing 
in those systems; we call the resulting state 'topological spin glass', as 
spin freezing emerges from  a substrate of a topological spin liquid, in our example the topological 
Coulomb phase of spin ice. 

\begin{figure}
{\includegraphics[width=\hsize]{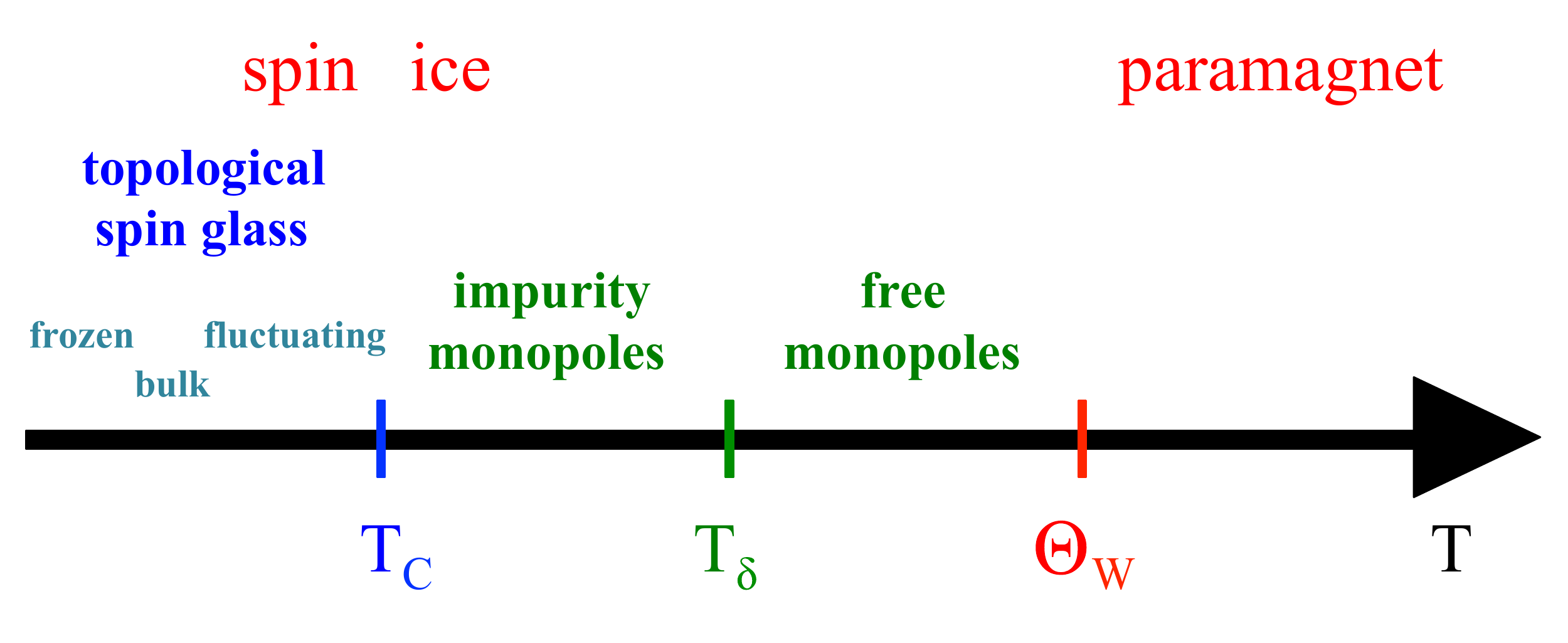}} 
\caption{Phase diagram of spin ice with a small density $x$ of spins removed. 
The glass transition takes place at $T_c$, while the crossovers 
from the high-temperature paramagnet to the topological Coulomb phase, and to
impurity dominated monopole excitations take place at $\Theta_W$ and $T_\delta$, respectively. \label{dsi:pholia}}
\end{figure}

\unue{Phase diagram of diluted spin ice} Our central result is the phase diagram, Fig.~\ref{dsi:pholia}, which shows the onset of the topological Coulomb regime at a temperature set by the cost of a topological defect (magnetic monopole) $\Theta_W\sim\Delta$. At a lower temperature, T$_c$--proportional to the dilution (density of missing spins) $x$--there is a spin glass transition.  Below T$_c$, the correlations characteristic of the topological Coulomb phase persist alongside a small frozen moment. 

Remarkably, our theory is very simple when cast in terms of  {\em missing} spins, 
which we call ghost spins. This occurs much in the same way that an almost  
filled band of electrons is most simply described in terms of dilute, positively charged 
holes, i.e.\ {\it missing} electrons. 

\begin{figure}
{\includegraphics[width=0.48 \columnwidth]{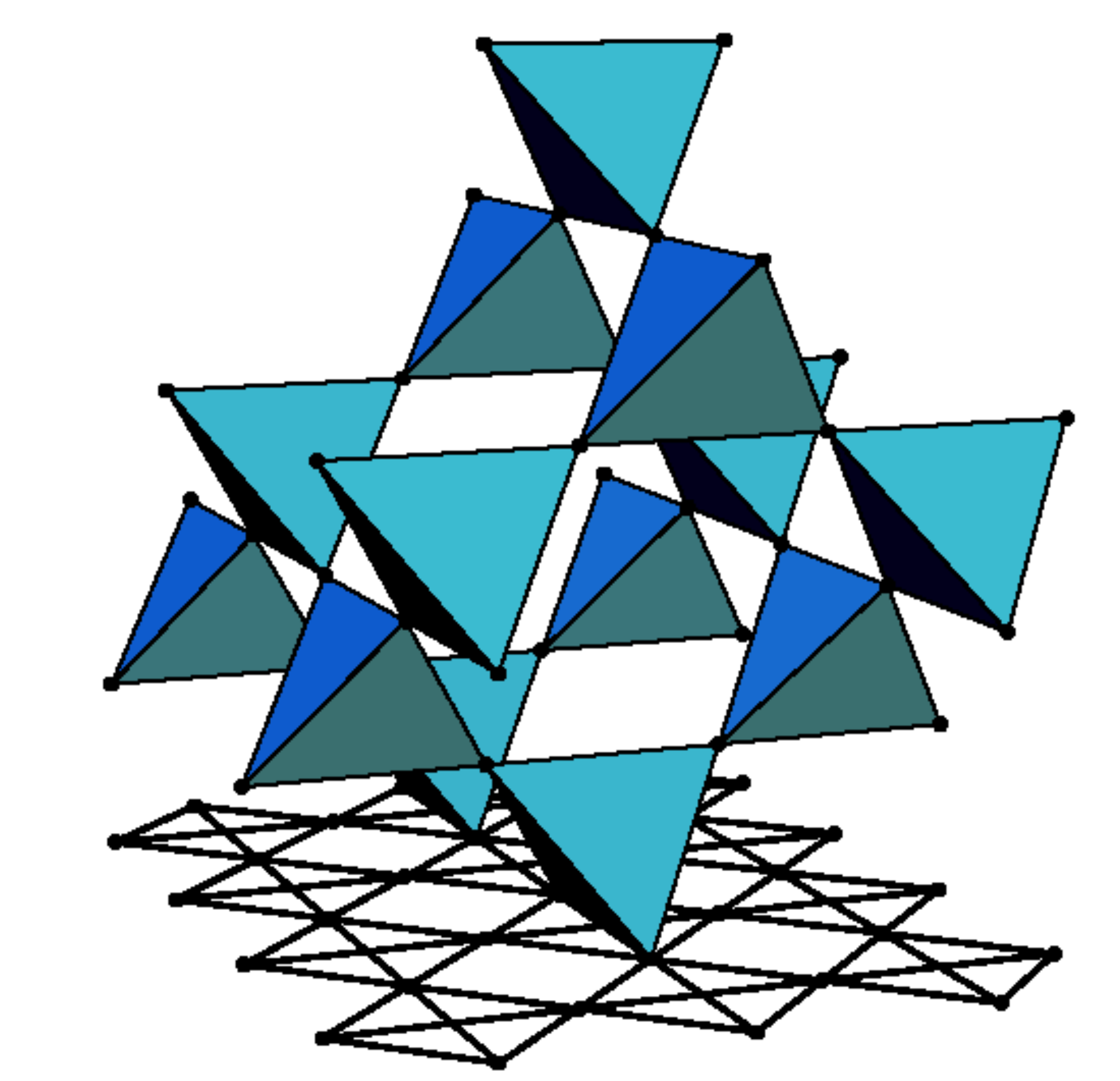}}%
{\includegraphics[width=0.48 \columnwidth]{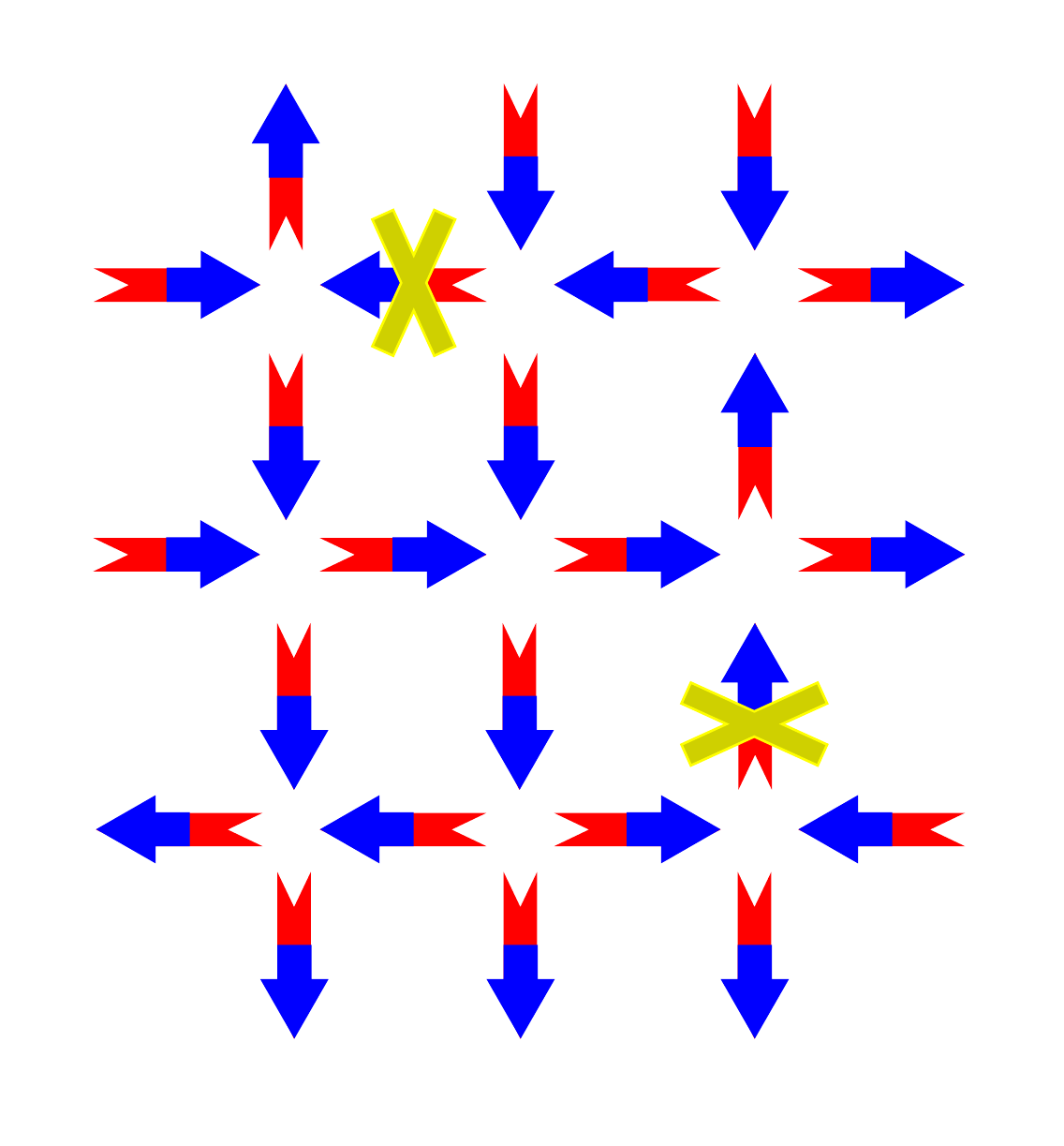}} 
{\includegraphics[width=0.48 \columnwidth]{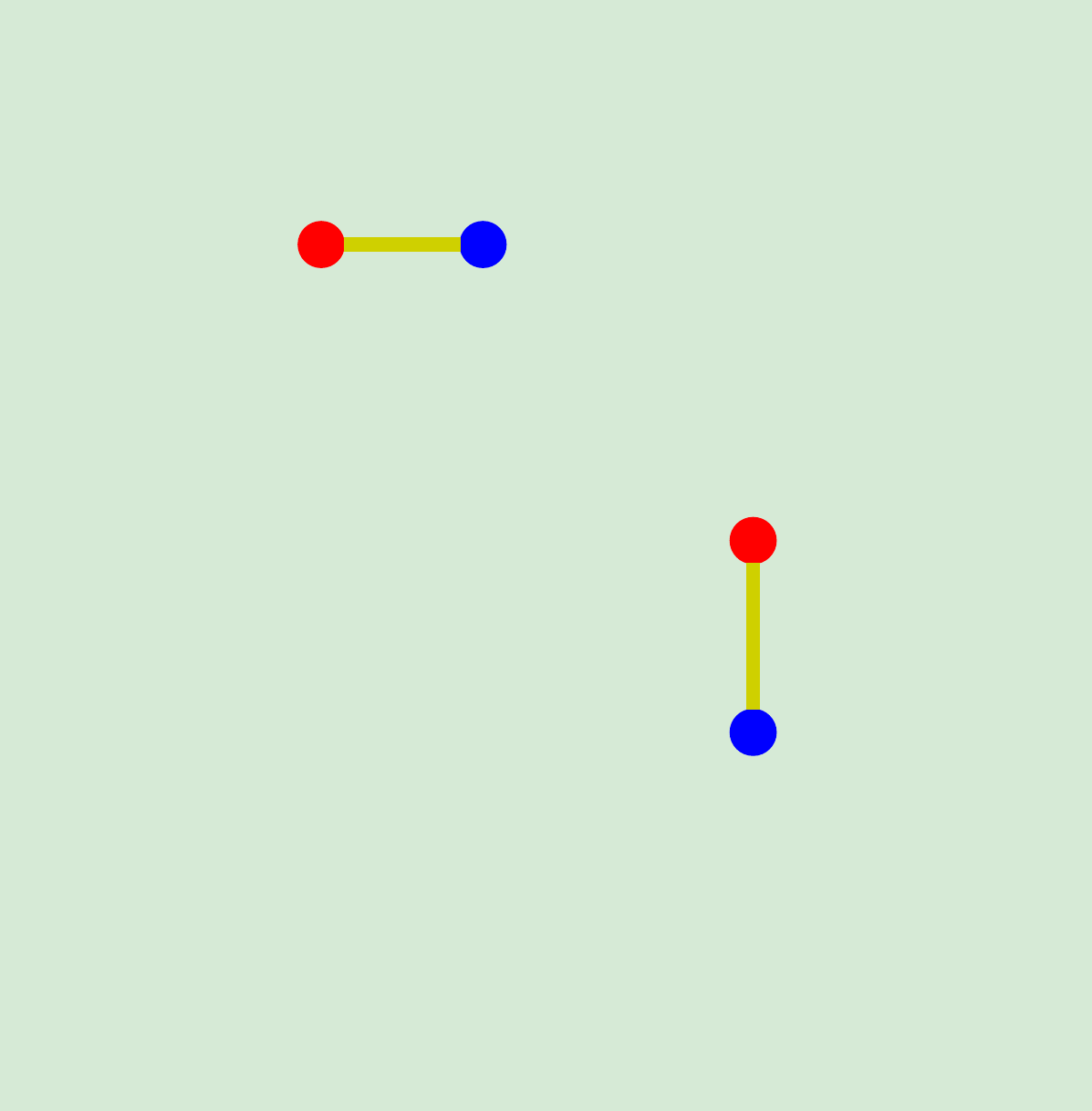}}%
{\includegraphics[width=0.48 \columnwidth]{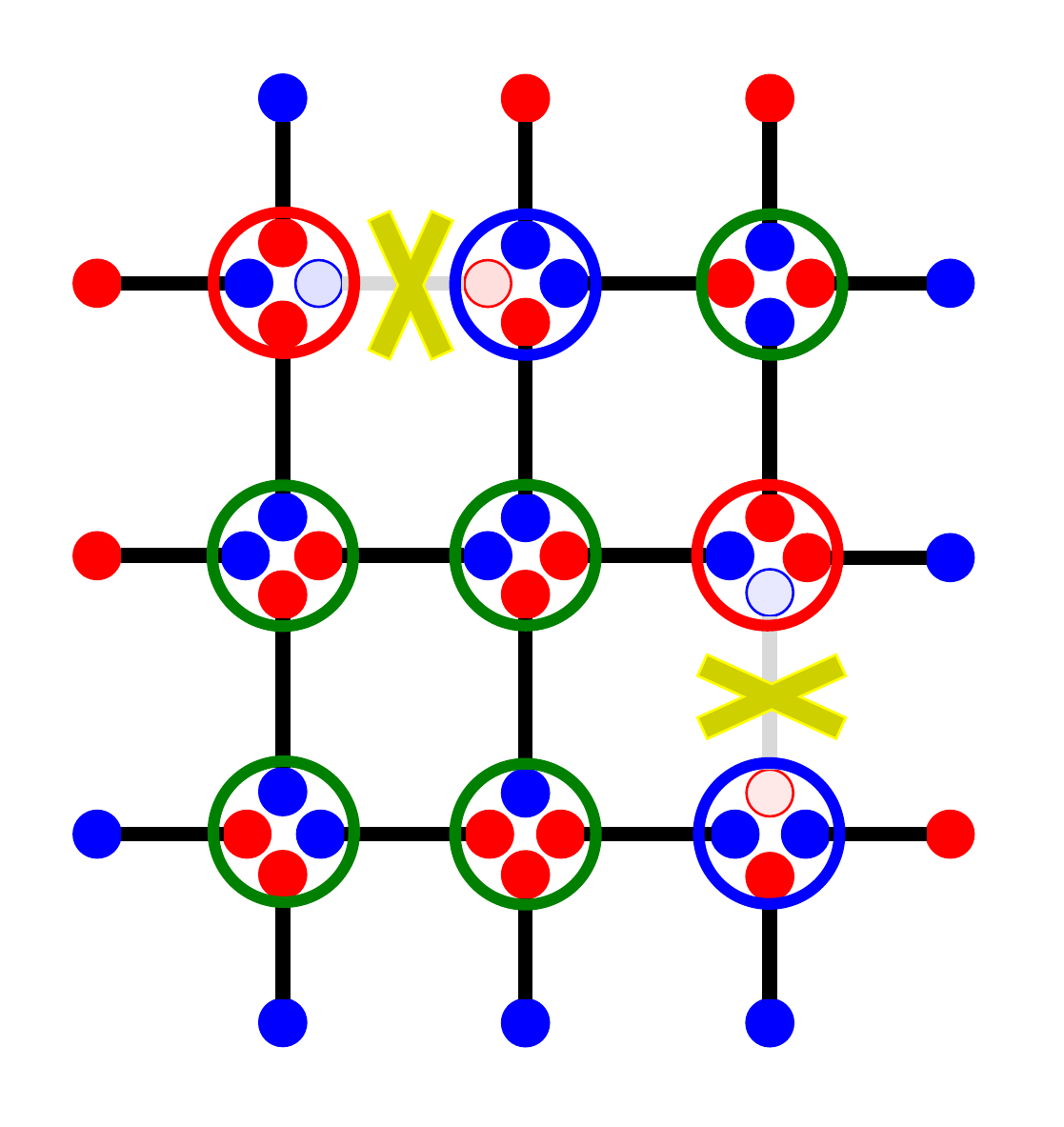}} 
\caption{Complexity reduction via genesis of ghost spins. The cartoons are for a projection of spin ice onto the two-dimensional plane (top left panel) for clarity; the diamond lattice defined by the centres of the tetrahedra thus turns into a square lattice. 
At low temperature $T \ll \delta, \Delta$, spin ice with a small density $x$ of {\em missing spins} (crossed out in top right) is equivalent 
 to a  small density of  {\em ghost spins} (bottom left). 
 This is straightforwardly established by writing each spin as a dumbbell of equal and opposite
 magnetic charges. 
At vertices either end of  a missing spin, the net charge is nonzero (red and blue circles), 
 so that they form the ends of the dipole of the ghost spin. At all other vertices, 
 the net charge vanishes (green circles) and can thus be omitted, so that  
 this pristine bulk of spin ice only provides an effective medium (shaded green) 
 carrying an interaction between the ghost spins, Eq.~\ref{eq:effD}
 and Fig.~\ref{dsi:entplusD}.}
\label{fig1}\label{doi:holespin}
\end{figure}

\unue{The dumbbell model and ghost spins} 
In dipolar spin ice, the degrees of freedom 
are Ising spins on the pyrochlore lattice (Fig.~\ref{doi:holespin}) whose local easy axis directions, 
$\hat{e}_i$, 
are defined by the line joining the centres of the pair of tetrahedra which share them; 
the simplest appropriate interaction Hamiltonian 
of Ising spins with moments $\vec\mu_{i,j}$ of size $\mu$, separated by $r_{ij}$,
 contains  short-range exchange interactions in 
addition to the usual magnetic dipolar term, $D{\cal D}_{ij}$, of strength $D$, with
\begin{equation}
{\cal D}_{ij}=
\frac1{\mu^2}\left( \frac{a}{r_{ij}} \right)^3 \left( \vec\mu_i\cdot\vec\mu_j - 3 ( \vec\mu_i\cdot\hat{r}_{ij})( \vec\mu_j\cdot\hat{r}_{ij})\right) 
\label{eq:effint}
\end{equation} 
with $a$ the nearest neighbour distance on the pyrochlore lattice, and $a_d=a\sqrt{3/2}$ the 
separation between the centres of the tetrahedra, which define a diamond lattice. 

To derive our central results, we appeal to the equivalent dumbbell 
model introduced in the prediction of the existence of 
magnetic monopoles.\cite{Nat451_2008}  
Here, each spin is represented by two equal and opposite magnetic charges ${\cal Q} = \mu / a_d$.
The only details of the model we require\cite{Nat451_2008} is that the pairwise spin interactions can be rewritten 
as a pairwise interaction of the total charges of each tetrahedron, ${\cal Q}_{tet}$. This includes  
a contact interaction $\Delta({a_d}/2{\mu})^2\sum_{tet} {\cal Q}_{tet}^2$ to reproduce the nearest neighbour interaction correctly,  in addition to 
standard (magnetic) Coulomb interactions between any other pair of 
charges. From these and the nearest neighbour exchange $J$, we can construct two energy scales; $D=\mu_0 \mu^2/(4\pi a^3)$, the coupling constant of the dipolar interaction (with vacuum permeability $\mu_0$) and $\Delta= \frac{2J}{3}+\frac{8}{3}\left(1+ \sqrt{\frac{2}{3}} \right)D$, the energy cost of a $|{\cal Q}_{tet}|=2{\cal Q}\neq0$ defect. 

Crucially, any of the exponentially many configurations satisfying the ice rule -- that  two spins point into each tetrahedron and 
two out\cite{Sc294_2001} --  is an exact ground state of this model.  This amounts to each diamond lattice site being charge neutral: {\em in any ground state, 
the charge density vanishes locally} \cite{Nat451_2008}.

This is the basis for what follows. Removing a spin by chemical substitution of a magnetic by 
a non-magnetic ion,  leaves behind two adjacent tetrahedra with equal and {\it opposite} 
charges $\pm {\cal Q}$,  a dipole $-\vec{\mu}$  -- the ghost spin --  with moment
opposite to that of the removed spin. Just like the charge 
of a hole in a semiconductor being the opposite of that of the missing electron, here it is the spin's magnetic moment
 which has changed sign.

The effective energetics follows simply by keeping track of the interactions between those charges -- the 
pairwise interaction between separated ghost spins, $\tilde{\cal H}_{ij}$ has the standard dipolar form
(again in complete analogy to the Coulomb repulsion between holes in a semiconductor):
\begin{equation}
\tilde{{\cal H}}_{ij}=\tilde{D}{\cal D}_{ij}
\label{eq:effint}
\end{equation}

Note the tremendous complexity reduction -- a dense, weakly diluted system of dipolar spins is described in terms of a 
low density of ghost spins! However, the intricate nature of the spin ice phase has not vanished entirely. The following are its most striking 
manifestations.

Firstly, the fractionalised excitations of the topological spin ice phase can be nucleated at the missing spin. In detail,  
a pair of ghost spins can be turned into a pair of impurity magnetic monopoles by flipping a string of spins
arranged head-to-tail
running between them.  For well-separated monopoles,
the resulting configuration  is higher in energy by 
$\delta = {4\sqrt{2}D}/{(3\sqrt{3})}$ per monopole. 

\begin{figure}
{\includegraphics[width=\hsize]{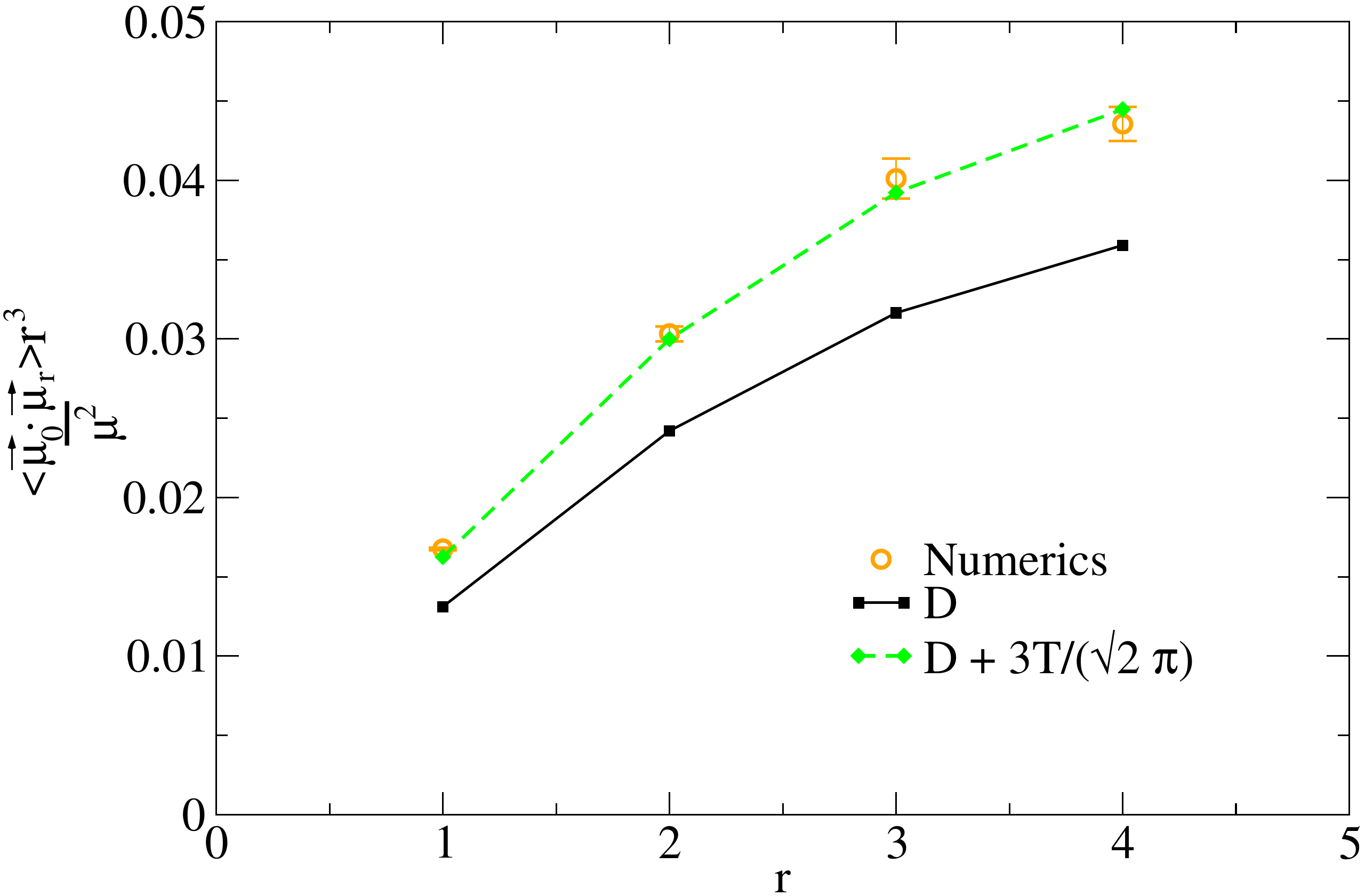}} 
\caption{Correlations between a pair of ghost spins $\vec{\mu}_0$ and $\vec{\mu}_r$ in an otherwise fully populated sample of spin ice (of linear dimension $L=12$, with $D=1.41$ K and $T=0.5$ K), 
separated by $r$ units in the $[101]$ direction. This is compared to 
two isolated magnetic dipoles in an otherwise empty unit cell;
agreement with numerics is only achieved upon including 
the entropic interactions mediated by the spin ice bulk, for which 
 the dashed line denotes the asymptotic form, Eq.~\ref{eq:effD}.\label{dsi:entplusD}}
\end{figure}

Secondly, the ghost spins do not disrupt the correlations in spin ice, as they have zero net magnetic 
charge, so  that in particular the pinch-points\cite{isakovdippyro, PRL110_2013,fennell_sci} found in neutron scattering, 
which reflect the emergent gauge field defining the Coulomb phase, remain intact. 

Thirdly, the effective dipolar coupling constant between the ghost spins $\tilde{D}$, Eq.~\ref{eq:effint},
has a contribution coming from the fluctuations of the spins in the bulk on top of the simple magnetostatic 
coupling constant $D$:
\begin{equation}
\tilde{D}=D + \frac{3T}{\sqrt{2} \pi}
\label{eq:effD}
\end{equation}
This happens because the number of spin ice states compatible with a given configuration of ghost spins depends 
on their relative orientation; this yields an entropic contribution, $J^\mathrm{ent}$, to the 
spin interaction\cite{isakovdippyro} -- 
for details, see the Suppl.\ Mat., where we derive the expression
\begin{equation}
{J_{ij}^\mathrm{ent}} = -T \langle \sigma_i\sigma_j\rangle_{\mathrm{nn}} ,
\end{equation}
where $\langle \sigma_i\sigma_j\rangle_{\mathrm{nn}}$  stands for the correlations between Ising spins on 
sites $i$ and $j$
in spin ice with the dipolar interactions switched off entirely; this gives Eq.~\ref{eq:effD} for $r_{ij}\gg a_d$
but is accurate also for small $r_{ij}$ (Fig.~\ref{dsi:entplusD}).

In other words, thanks to the Coulomb phase, the missing
dipolar spins know about the correlations they would have if they were neither missing
nor dipolar!

\unue{Freezing of the ghost spins} 
It is notoriously hard to simulate spin freezing transitions~\cite{sim1,sim2,sim3},
all the more so in this case where a small number of 
ghost spins requires simulation of a much larger number of bulk spins. However, thanks to the abovementioned complexity 
reduction, this effective problem, Eq.~\ref{eq:effint}, can be analysed, still  with a considerable amount
of numerical effort. 

\begin{figure}
{\includegraphics[width=\hsize]{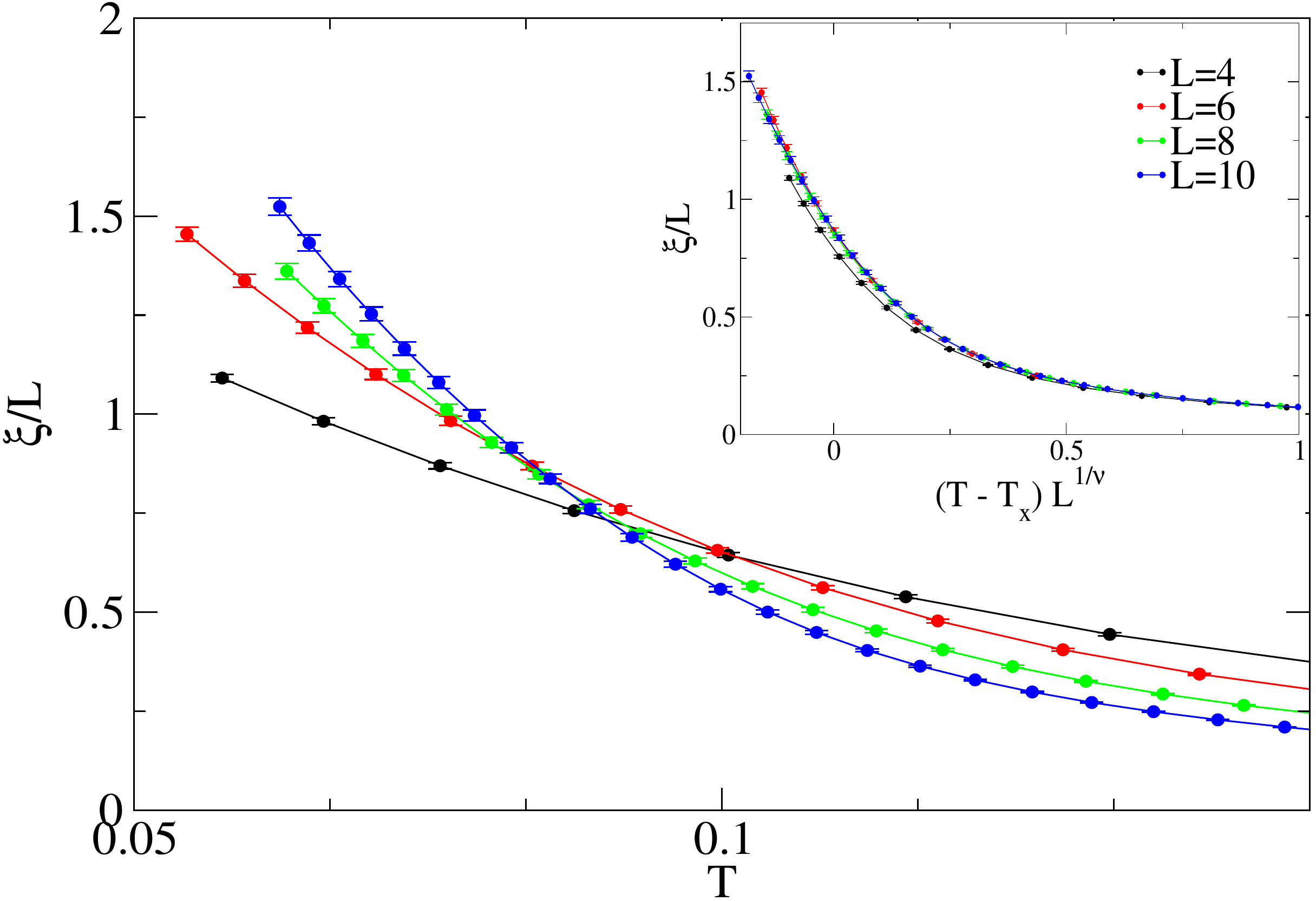}} 
\caption{The freezing transition. Number of spins equals $16L^3x$ in a system with $L^3$ unit cells, 
with $x=0.0625$. Dipolar coupling of $D=1.41$ K (as in Ho$_2$Ti$_2$O$_7$ and Dy$_2$Ti$_2$O$_7$)\cite{Sc294_2001}
is used in simulations: the crossing point in $\xi/L$ (see Suppl. Mat.)
indicates a continuous glass transition, further evidence for which is the  data collapse (inset) for different system sizes, for 
$\nu=1.08(3)$ and critical temperature $T_x = 0.084(2)$.}
\label{numericsdata}
\end{figure}

We have demonstrated numerically that there is spin freezing for the random dipolar model, Fig.~\ref{numericsdata}.
We  find a critical temperature T$_x \propto x$, as one would expect for a dipolar system 
with typical distance between  spins $r\sim x^{-1/3}$, and hence  dipolar interaction energy
scale $\sim r^{-3}\sim x$.  We find numerically that T$_x \simeq 0.95 D x$, which 
implies an entropically renormalised value of $T_c(x)=T_x/(1-\frac{3T_x}{\sqrt{2}\pi D})$ for the freezing transition into the 
topological spin glass. 

This  complements known cases of freezing for random dipoles, namely dense dipoles on a cubic lattice 
with random orientations;\cite{denseglass} or dilute but collinear dipoles on a cubic lattice.\cite{gingras}
In our case, the dipoles are dilute and their orientations are neither random nor collinear, 
being picked  from the local easy axes of the occupied sites, respecting the cubic symmetry of the pyrochlore lattice.

\unue{Energy scales and the role of perturbations}
Let us now examine the phase diagram Fig.~\ref{dsi:pholia} in more detail. At high temperature, T$\gg\Delta$, we have an ordinary disordered paramagnet. Below $\Theta_W\sim\Delta$, the ice rules are enforced, yielding the topological Coulomb regime with experimentally sparse monopoles as elementary excitations, $\rho \sim e^{-\Delta/T}$.\cite{Nat451_2008} 
However, the impurity magnetic monopoles mentioned above have a cheaper energy cost $\delta<\Delta$ 
than a monopole in the bulk as no energy needs to be paid for violating the ice rule in the first place 
-- this is taken care of by the quenched chemical dilution. The corresponding density is $\rho_x \sim xe^{-\delta/T}$, 
which dominates over $\rho$ at lower temperatures below $T_\delta = (\delta-\Delta)/\ln x$. Further, at T$_c$, the glass transition occurs, at which (presumably\cite{althou,katzyoung}) Ising symmetry is broken. The bulk spins in the Coulomb phase continue to fluctuate, however. 

This is the above-mentioned  coexistence of the topological phase and the spin glass phase for which it provides 
the substrate--the topological spin glass.

The frozen moment -- the 'order parameter' $q_{EA}=\frac{1}{N}\sum_{i=1}^N\langle S_i\rangle^2$ of the spin glass -- 
 appears at  $T_c$ and
 grows as the temperature is lowered, as do static local fields set up by the frozen
moments $B_f \sim (\mu_0\mu x\sqrt{q_{EA}})/(4\pi a^3) $ \cite{intfields}. The resultant Zeeman energy 
$\sim D\sqrt{q_{EA}}x$ will try to pin the bulk spins to point along the direction of the local fields,
against the entropy of fluctuations between different spin ice configurations,\cite{aprbss} estimated  by Pauling to be  
$S_p \approx \frac{1}{2}\ln \frac{3}{2}$ per spin.
The glass transition at $T_c$ being continuous, for $T\alt T_c$ the frozen moment and 
concomitant Zeeman energies will be very small, and
the bulk spins will continue to fluctuate essentially like in pure spin ice. As $T$ is lowered further,
the entropic contribution of the bulk fluctuations to the free energy vanishes approximately $\propto T$, 
while $q_{EA}$ grows, so that the system 
eventually freezes into one spin ice configuration. Since the static fields are too weak to break the ice rules,
the frozen state still exhibits the correlations of the Coulomb phase
when averaged over the entire sample \cite{AnnRCMP3_2012,Sc294_2001}.

\unue{Comments on experiment and outlook}
Generally speaking, 
the low-temperature physics of frustrated systems is non-universal, and the 
topological spin glass may be preempted by perturbations to the dumbbell Hamiltonian if they are large enough. 
Indeed, the question what happens in diluted spin ice compounds \cite{PRL99_2007} was recently addressed in 
detailed numerical simulations,\cite{arXiv13037240} which emphasized the need to consider the complete 
Hamiltonian to obtain a fit of theory to experiment. The question of whether freezing generically
occurs was not settled there. 

The central point of our work is that there exists a {\em microscopic} Hamiltonian for which  
the existence of the topological spin glass can be predicted 
with a high degree of confidence. Features such as the simultaneous
appearance of topologically spin liquidity and glassiness appear naturally, along with the presence of a small frozen moment
alongside a sizeable fluctuating component. The challenge for establishing
its existence in experiment is thence to find a compound avoiding other instabilities both of the non-dynamic 
and dynamic nature, such as a non-cooperative slowing down.\cite{RyzPet_2005,JauHol_2009} This seems 
a very
realistic prospect given the wide range of spin ices available nowadays, with the combination alone
of A$_2$B$_2$O$_7$, (A=Dy, Ho, Tb, Yb, $\ldots$, B=Ti, Ge, Sn, $\ldots$) providing numerous 
examples differing in many fundamental parameters such as the relative size of exchange and dipolar interactions, 
as well as many single-ion properties.\cite{gingclar}

More broadly, we would like to emphasize the {genericity} of the ingredients involved in our study. 
We used the ``vacuum'' of a Coulomb phase and its local charge neutrality, and the fact that defects therein 
will  have an interaction determined by the emergent gauge theory describing the low 
energy physics of the topological phase. The detailed resulting collective behaviour will be as varied as the 
richness of the latter ingredients; therefore, it is clear that an outcome in which randomly located emergent degrees of freedom interact 
via highly frustrated interactions is a generic one, and so is the expectation of a topological spin glass.

{\it Acknowledgements:} We are very grateful to Michel Gingras, Shivaji Sondhi and Claudio Castelnovo for useful discussions, to the latter two for collaboration on related work, and to the last for detailed comments on the manuscript.

\appendix

\section{Derivation of the effective Hamiltonian}

We first establish Eq.~\ref{eq:effD}, that the ghost dipoles experience an enhanced
interaction compared to free dipolar spins, mediated by the spin ice bulk. This form 
is valid at low temperature, where monopoles are exponentially sparse so that they can be neglected
to an excellent approximation. 
Our demonstration consists of 
showing that the dipolar and entropic part of the interactions simply add; and that the latter
is reflected in the correlations of spins in 'nearest-neighbour' spin ice, i.e.\ 
in the absence of magnetic dipolar interactions and disorder. 

For this case, we denote the Ising spins with variables $\sigma_i$. It is  a non-trivial but known fact
that the correlations induced by the entropic interactions are dipolar,\cite{isakovdippyro} with
\begin{equation}
\langle \sigma_i \sigma_j \rangle_{T=0} = -\frac{3}{\sqrt{2} \pi} \left(\frac{a}{r_{ij}} \right)^3 (\hat{e}_i \cdot \hat{e}_j - 3(\hat{e}_i \cdot \hat{r}_{ij})(\hat{e}_j \cdot \hat{r}_{ij})) \ ,
\end{equation}
where the numerical prefactor
${3}/{\sqrt{2} \pi}$ is  computed approximately but accurately from a large-$n$ ansatz. 

This result is reproduced by a simple pairwise effective free energy, which keeps track of the relative number
of spin ice states compatible with the four possible configuration of the spin pair:
\begin{eqnarray}\label{fent}
\mathcal{F}_{\sigma_i\sigma_j}& = &J^\mathrm{ent}_{i,j}T\sigma_i\sigma_j \Rightarrow -T \langle\sigma_i\sigma_j\rangle_{T=0}=J^\mathrm{ent}_{i,j} \ , 
\end{eqnarray}
which follows upon linearising the Boltzmann factor $\exp(-\mathcal{F}/T)$, which is appropriate 
as these correlations are small. These equations together establish an entropic coupling constant ${3T}/{\sqrt{2}\pi}$, which
as we show next is the difference between  $\tilde{D}$ and $D$. 

Let us now consider the full dipolar interaction problem and reinsert a dipole each at the location of 
the two ghost spins as to obtain 
a state obeying the ice rules. In the
absence of monopoles at low temperature, this is always possible, crucially,  in a {\em unique} way. 
This means that the
counting of the number of configurations compatible with a given state of the ghost spins does not
depend on their presence or absence, so that the (unnormalised) Boltzmann factor for a given ghost
spin configuration is simply the multiplicity -- given by 
the entropic free energy, Eq.~\ref{fent} -- multiplied by the energetic Boltzmann factor due to the dipolar interaction. 
Together with Eq.~\ref{eq:effint}, we thus obtain the full interaction between the ghost spins as
\begin{equation}
\left(D+\frac{3T}{\sqrt{2}\pi} \right) \left(\frac{a}{\mu r_{ij}} \right)^3 (\hat{\mu}_i \cdot \hat{\mu}_j -3 (\hat{\mu}_i \cdot \hat{r}_{ij})(\hat{\mu}_j \cdot \hat{r}_{ij})) \ ,
\end{equation}
(using $\hat\mu_i\parallel\hat{e}_i$ and $\vec\mu_i=\mu\hat\mu_i$); this gives Eq.~\ref{eq:effD}.

\section{Freezing transition at small dilution}
We use Monte-Carlo simulations to study the freezing transition in very dilute spin ice, obtained after mapping the weakly diluted system of dipolar spins to a low density of ghost spins, and show that there is a continuous transition to a spin glass phase. A single-spin flip Metropolis algorithm in combination with parallel tempering in temperature\cite{sim1} is used to ensure proper equilibration in the glassy phase. The data is averaged over several disorder realizations (typically $1000$ or more), each of which is produced by placing spins on a fraction $x$ of sites that are randomly selected from a total of $16L^3$ sites in a system of linear size $L$ (with $16$ sites in the conventional cubic unit cell for the pyrochlore lattice). The long-ranged nature of the dipolar interactions is treated using the Ewald summation technique (e.g. see Z.Wang and C. Holm, J. Chem. Phys. {\bf 115}, 6351 (2001)).

The spin glass order parameter, $q_{EA}^{\alpha \beta}(\mathbf{k})$, at wavevector $\mathbf{k}$ is defined as
\be
q_{EA}^{\alpha \beta}(\mathbf{k}) = \frac{1}{N}\sum_i \mu_i^{\alpha (1)} \mu_i^{\beta (2) }\exp(i\mathbf{k} \cdot \mathbf{r}_i) 
\ee
where $N=16L^3x$; $\alpha,\beta=x,y,z$ are the spin components (where the ghost spins point along the local easy axes) and $(1)$ and $(2)$ denote two identical disorder realizations of the system with the same set of interactions. From this, we calculate the spin glass susceptibility, $\chi_{SG}(\mathbf{k})$, defined as
\be
\chi_{SG}(\mathbf{k}) = N \sum_{\alpha,\beta}[\langle|q_{EA}^{\alpha \beta}(\mathbf{k})|^2 \rangle] 
\ee
where $\langle \cdots \rangle$ and $[\cdots]$ denote thermal and disorder averages, respectively.

A spin glass correlation length $\xi$ can then be defined by using the following relation:
\be
 \xi=\frac{1}{2\sin \left(\frac{|\mathbf{k}_{\rm min}|}{2} \right)} \left(\frac{\chi(0)}{\chi(\mathbf{k}_{\rm min})}-1 \right)^{\frac{1}{2}}
\ee
where $\mathbf{k}_{\rm min} = \frac{2\pi}{L}(1,0,0)$. If the freezing transition is continuous, then $\xi$ is expected to satisfy a scaling relation of the form $\xi/L = \mathcal{F}((T-T_x)L^{1/\nu})$ where $\mathcal{F}$ is a universal scaling function and $T_x$ is the critical temperature. From this, it follows that $\xi/L$ for different sizes $L$ (for sufficiently large $L$) should cross at $T_x$. The behaviour of $\xi/L$ from the numerics for small $x$ indeed confirms to this expectation (see Fig~\ref{numericsdata}) and provides numerical evidence for a continuous spin glass transition in a dilute dipolar spin ice. 

\begin{figure}
{\includegraphics[width=\hsize]{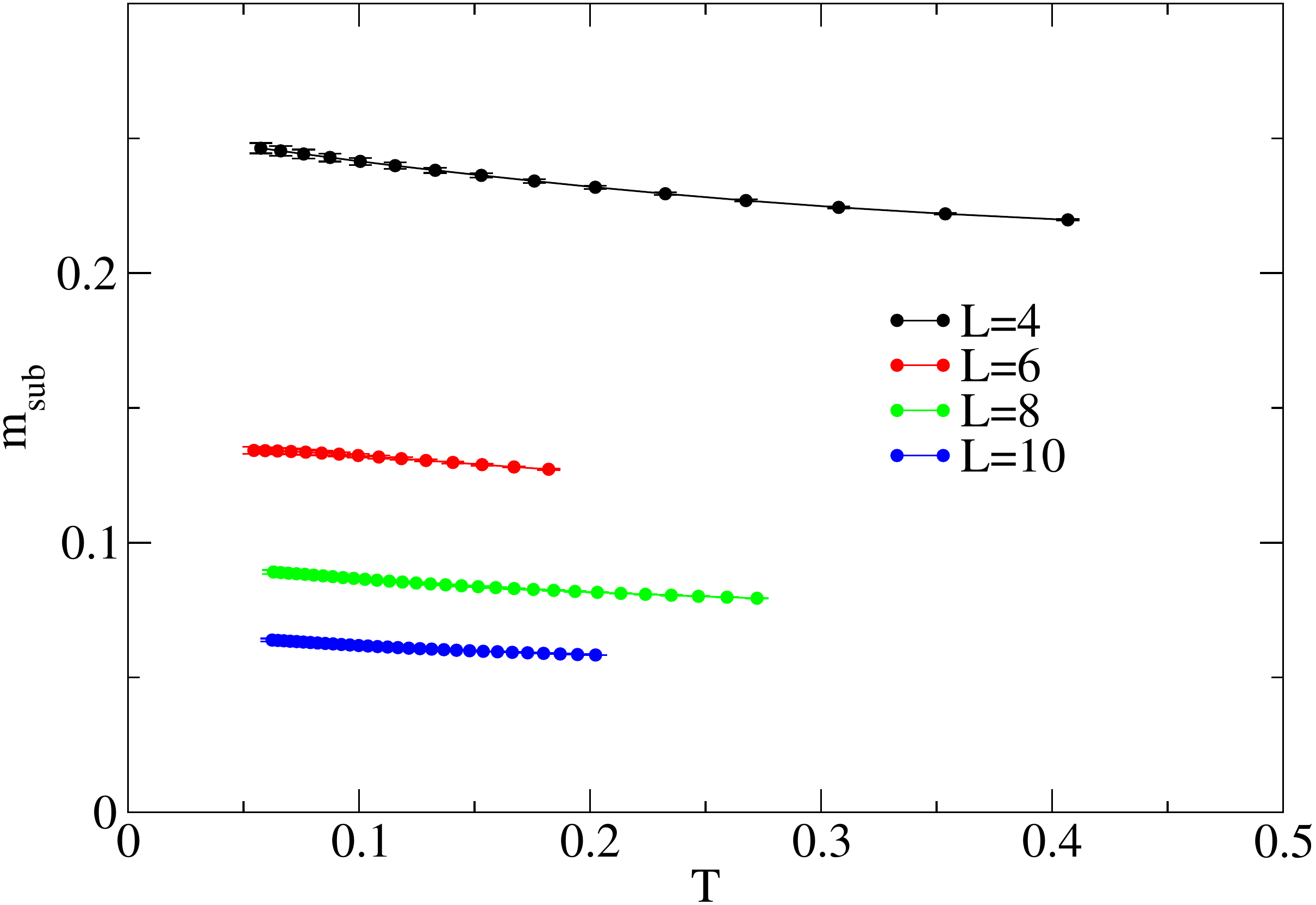}} 
\caption{The behaviour of sublattice magnetization $m_{sub}$ for $x=0.0625$ at coupling $D=1.41$ K. The data shows the absence of any long-range ordering of $m_{sub}$ on lowering the temperature below the glass transition $T_x$.}
\label{sublattice}
\end{figure}

We have also checked the behaviour of other quantities to rule out any obvious long-ranged ordered states. For example, we evaluate the sublattice magnetization per site, $m_{sub}$, defined as
\be
m_{sub}=[\langle|\frac{1}{N_a}\sum_i S_{i,a}|\rangle] 
\ee
where the sum runs only over sites that belong to the same sublattice $a$ (the pyrochlore lattice is a FCC lattice with a four-point basis), $N_a$ counts the sites in sublattice $a$ for a given disorder realization and $S_i=(\vec{\mu}_i \cdot \hat{e}_i)/\mu=\pm 1$. In Fig. \ref{sublattice}, we see that $m_{sub}$ scales as $L^{-3/2}$ with increasing $L$, indicating the absence of any ordering transition to a state with non-zero sublattice magnetization.     

\end{document}